\documentclass[10pt,letterpaper]{article}
\usepackage[top=0.85in,left=2.75in,footskip=0.75in]{geometry}

\usepackage{amsmath,amssymb}

\usepackage{changepage}

\usepackage[utf8x]{inputenc}

\usepackage{textcomp,marvosym}

\usepackage{cite}

\usepackage{nameref,hyperref}

\usepackage[right]{lineno}

\usepackage{microtype}
\DisableLigatures[f]{encoding = *, family = * }

\usepackage[table]{xcolor}

\usepackage{array}

\usepackage{subcaption}

\newcolumntype{+}{!{\vrule width 2pt}}

\newlength\savedwidth

\raggedright
\usepackage[aboveskip=1pt,labelfont=bf,labelsep=period,justification=raggedright,singlelinecheck=off]{caption}

\makeatletter
\renewcommand{\@biblabel}[1]{\quad#1.}
\makeatother

\usepackage{lastpage,fancyhdr,graphicx}
\usepackage{epstopdf}

\usepackage{caption,subcaption}
\usepackage{multirow}
\usepackage{enumitem}
\usepackage{hyperref}

\usepackage{xcolor}

\fancyheadoffset[L]{2.25in}
\fancyfootoffset[L]{2.25in}
\begin{document}
\vspace*{0.2in}

\begin{flushleft}
{\Large
\textbf\newline{Streetonomics: Quantifying Culture Using Street Names} 
}
\newline
\\
Melanie Bancilhon\textsuperscript{1},
Marios Constantinides\textsuperscript{2*},
Edyta Paulina Bogucka\textsuperscript{4},
Luca Maria Aiello\textsuperscript{2,5},
Daniele Quercia\textsuperscript{2,3}
\\
\bigskip
\textbf{1} Washington University in Saint Louis, Saint Louis, USA
\\
\textbf{2} Nokia Bell Labs, Cambridge, UK
\\
\textbf{3} CUSP, King’s College London, London, UK
\\
\textbf{4} Technical University of Munich, Munich, Germany
\\
\textbf{5} IT University, Copenhagen, DK
\bigskip

%
%





* marios.constantinides@nokia-bell-labs.com

\end{flushleft}
\section*{Abstract}
Quantifying a society's value system is important because it suggests what people deeply care about---it reflects who they actually are and, more importantly, who they will like to be. This cultural quantification has been typically done by studying literary production. However, a society's value system might well be implicitly quantified based on the decisions that people took in the past and that were mediated by what they care about. It turns out that one class of these decisions is visible in ordinary settings: it is visible \emph{in street names}. We  studied the names of 4,932 honorific streets in the cities of Paris, Vienna, London and New York. We chose these four cities because they were important centers of cultural influence for the Western world in the 20\textsuperscript{th} century. We found that street names greatly reflect the extent to which a society is gender biased, which professions are considered elite ones, and the extent to which a city is influenced by the rest of the world. This way of quantifying a society’s value system promises to inform new methodologies in Digital Humanities; makes it possible for municipalities to reflect on their past to inform their future; and informs the design of everyday’s educational tools that promote historical awareness in a playful way.



\section*{Introduction}
\label{sec:section1}

Culture is a complex multi-level construct~\cite{taras2009half}. Anthropologists claim that it is intangible and imperceptible due to our inability to quantify it~\cite{white1959concept}. In this study, we define a society's culture as its set of collective behaviors, norms and values.

To study geographical variations of cultural aspects, traditional approaches have resorted to surveys, census, or literary production. More recently, however, the availability of online open-data sources has made it possible to study and quantify cultural aspects on a larger scale. Michel et al.~\cite{michel2011quantitative} introduced `culturomics': a large-scale study of millions of digitized books that uncovered historical trends. Montalto et al.~\cite{montalto2019culture} quantified a city's culture by investigating indicators such as cultural vibrancy (e.g., presence and attractiveness of cultural venues and facilities), creative economy (e.g., the capacity of culture to generate jobs and innovation), and creative potential (e.g., conditions that enable creative processes to thrive), and they did so for multiple European cities.

We connect these two veins of research by proposing \emph{`streetonomics'} (\url{http://social-dynamics.net/streetonomics}), an alternative way of quantifying cultural indicators using a type of urban features that is visible yet often overlooked: \textit{street names}. ``The main merit of commemorative street names is that they introduce an authorized version of history into ordinary settings of everyday life'', as Azaryahu~\cite{azaryahu1996power} put it. Street names are more than spatial indicators, and, since the beginning of time, rulers have used spatial engineering as a form of social engineering. As a result, street names mirror a city's social, cultural, political, and even religious values~\cite{oto2017street}. 

For the first time, we combined heterogeneous open-data sources to offer a unique and cheap way to study the footprint of a city's cultural dimension through time and space. This approach could offer tools that would not only create dwellers' awareness of the `historical memory' of a city, but would also be used by public authorities to reflect on past street-naming decisions and to inform future ones. In doing so, we made three main contributions:

\begin{itemize}
\item We formulated four main research questions (\S\ref{sec:section3}3) that are meant to quantify: \emph{i)} gender biases in naming streets; \emph{ii)} whether streets speak to a distant past \emph{vs.} a closer present; \emph{iii)} which professions have been historically celebrated and which eventually died off; and \emph{iv)} whether culture historically had a local \emph{vs.} global focus.

\item We curated and made publicly available a dataset containing street names and information about the figures---`honorees'---these streets were named after (\S\ref{sec:section4}4). The dataset contains 4,932 streets in the cities of Paris, Vienna, London, and New York.
\item As we hypothesized, we found that street names reflect the value system of a society (\S\ref{sec:section5}5). Street names are gender-biased against women in two cities: the highest female proportion is found in Vienna (54\%), followed by London (40\%), then Paris (32\%), and then New York (26\%). Streets speak to a distant past, all the more so for those in London, Vienna and in Paris rather than those in New York. Certain professions have been consistently celebrated over centuries (e.g., artists, scientists, and writers), while others  went out of fashion (e.g., military). Finally, based on the number of streets named after foreigners, among the four cities, Vienna was the city that celebrated nations other than its own.

\end{itemize}

\section*{Related work} 
\label{sec:section2}

\subsection*{Urban Informatics} \label{subsec:section2_1}
By 2025, we will see another 1.2 billion people living in cities, and that will require new ways of managing the increased complexity that comes with larger cities. Since culture is one of the most important factors for urban success~\cite{florida2002rise}, we need new ways of `quantifying' it. Recently, Hristova et al.~\cite{hristova2018} quantified the cultural capital of neighborhoods in London and New York. They did so by operationalizing sociologist Pierre Bourdieu's theory of cultural capital upon geo-referenced pictures. They found that neighborhood's cultural capital  rather than its economic capital is more predictive of future house prices and, alas, of gentrification as well. By going beyond culture, researchers have recently quantified important sociological aspects of city life: from Jane Jacobs's urban vitality ~\cite{de2016death} to Stanley Milgram's urban ambiance~\cite{redi2018spirit}.

In line with this research, we propose the use of open data to quantify a city's culture, contributing to the fields of social computing and computational social science, in that, we answered questions typical of studies concerning culture or urbanism, and did so in a computational way.

\subsection*{Streets and Place Naming} \label{subsec:section2_2}
Naming streets and places is linked to power. Cohen and Kliot's~\cite{cohen1992place}  analyzed how place naming functioned as a strategy for governments to promote specific historical events over others and, in so doing, shape national identity. Guyot et al.~\cite{guyot2007identity} described place naming as a territorialization process that shapes place identify.  In a similar way, Azaryahu~\cite{azaryahu1996power} described street naming as an administrative control process. To see this type of processes in practice, consider  South Africa during the apartheid period. Place names originated by the fusion of European and Afrikaans names and served colonial powers in their attempt to erase indigenous groups' identities. To partly counter that, the new territorial demarcations have sought to address the inequalities stemming from the apartheid and shape a new national identity~\cite{guyot2007identity, azaryahu2002mapping}. In a similar way, in the United States, studies have shown that streets named after Martin Luther King were placed on minor streets rather than major roads~\cite{mitchell2014street}, thus reinforcing traditional racial
boundaries~\cite{rose2010geographies}, and recent street (re)naming has tried to change the country's course. Oto-Peralias~\cite{oto2017street} analyzed the religiosity expressed by street names in Spain, and found that it was correlated with cultural and economic factors. Contrary to this previous work, our study differs in two ways: \emph{i)} we analyzed a large corpus of streets comprised of four cities as opposed to a single city; and \emph{ii)} we explored honorees' `meta' information such as their genders, professions, countries of origin as opposed to studying naming patterns.

Overall, previous studies were limited in scale, or were restricted to the use of street names only. This study, instead, collated a large dataset of street names covering four large cities across two centuries, and relied on open-source knowledge bases to enriched these names with information about the historical figures after whom the streets were named, including age, gender, occupation, and country of origin.

\subsection*{Historical background of Paris, Vienna, London, and New York}

\mbox{ }\\
\noindent
\textit{\textbf{Paris}} The early changes in rational planning in Paris occurred in the Enlightenment period, where cemeteries were banned from the city due to the potential health hazard caused by the large number of corpses buried in churches during the French Revolution. Henri Lefebrve once remarked that ``new social relationships call for new space, and vice verca''. This is reflected in the decisions that the revolutionaries of 1789 made to establish the Conseil des Batiments Civils and built courts and prisons to reflect the tenets of the Age of Reason~\cite{knox2014atlas}. Later in 1860, Napoleon III decided to transform Paris into a modern empire by expanding the walls of the city and redesigning the urban scene. He hired urban planner Georges-Eugene Haussmann and together they started rebuilding and expanding the city. Between 1852 and 1870, the center was rebuilt with wide new avenues, parks and squares, and the city went from 12 to its current 20 ``arrondissements''. They built the luxurious Rue de Rivoli~\cite{knox2014atlas}, one of the most iconic streets in the city. As new streets were being built, a street naming convention was established to assign names to new streets and rename the existing ones.

\mbox{ }\\
\noindent
\textit{\textbf{Vienna}} In 1860, liberals in Austria took over the Habsburg Empire and transformed its institutions in accordance to the liberal identity and the cultural values of the middle class. In the process, they gained power over the city of Vienna. The city became a cultural and economic hub, and the center of intellectual life. From the moment of their accession to power, liberals began to reshape the city in their own image and transformed the face of Vienna. This era is commonly called the Ringstrasse era (``Ring Street Era''). This was the name of a new major road that was built to replace the old city walls and that separated the old city from its suburbs~\cite{schorske2012fin}. Since then, the city experienced a rapid expansion and the practice of naming new streets after historical figures started.

\mbox{ }\\
\noindent
\textit{\textbf{London}}
A defining moment for the British capital was 1666. During that time the Great Fire broke out, and the city of London was reborn from its ashes. The fire destroyed about 60\% of the city and, within a few days of the fire, plans for the rebuilding of the city commenced; these included restoration of churches, development of new streets and roads, and expansion~\cite{reddaway1951rebuilding}. In the following centuries, there was a period of rapid growth. The early signs of the Industrial Revolution were evident, and London became the centre of the evolving British Empire. Until 1920, it was the largest city in the world for a century.

\mbox{ }\\
\noindent
\textit{\textbf{New York}}
When the city of New York was still occupied by the Dutch, street names were primarily used to commemorate the city's European heritage. These names were later on Anglicized by the British: streets were renamed first after landowners to mark property ownership, and later after well-known figures, including merchants, war heroes and politicians. Despite the diffusion of this conventional naming practice, the municipality encountered issues establishing a complete system of street signs. As the city expanded, the Commissioners' Plan of 1811 was adopted, a grid urban planning was introduced, and streets were numbered rather than named. Only very few streets would be named after certain privileged people. Soon after, street co-naming was adopted in `honorific streets'~\cite{rose2008number}, where people's name were added to street numbers.

\section*{RESEARCH QUESTIONS}
\label{sec:section3}
As discussed in the previous section, we considered the four cities largely because, by looking at them, one looks at some of the Western world's most important cultural centers in the past century: in the first part of the 20\textsuperscript{th} century, Vienna was the cultural capital of the world with key figures such as Klimt and Freud; then, around the 1930s, Paris became the `go-to place' for American artists, and Hemingway and Gershwin were two notable examples; then, for a century until 1920s, London had become the largest city in the world; finally, in the 1950s, New York took center stage by redefining how art (Jackson Pollock) and literature (Jack Kerouac) were meant to look like in the modern world, and that influence still persists today. For those cities, we set out to answer four main research questions and, in so doing, we investigate whether explicit policies and public debates left any traces in our dataset.

\mbox{ }\\
\noindent
\textit{\textbf{Q1} Are streets names in Paris, Vienna, London, and New York gender-biased?}
Throughout history, western societies have often been masculinity-oriented, and women's achievements have been severely overlooked. For centuries, women were relegated to the private sphere, while men dominated the public sphere~\cite{eger2001women}. Urban planning was no exception; cities were built by men and for men---and that made street naming gender biased~\cite{roberts2016fair}. In Harlem, through the 1970s and 1980s, the names of African American women were removed from places of memory, which shows how commemorative strategies can reflect a combination of racial and gender discrimination~\cite{rose2017political}.

Gender equality in street names has been demanded by several feminist organizations around the world, including by the group called ``OsezLeFeminisme'' in Paris who protested that ``it is time to reinstate the place of women in history'' (\url{https://www.citylab.com/design/2015/08/the-streets-of-paris-renamed-for-women/402526}). As a result, in 2015, Paris Mayor Anne Hildalgo has renamed several streets after iconic women such as Nina Simone, Assia Djabar, Simone de Beauvoir, and Rosa Parks. 
We set out to explore whether these policies left any traces in our dataset.

\mbox{ }\\
\noindent
\textit{\textbf{Q2} Are the historical times in which the honorees lived close or distant from present times?}\\
The long/short term orientation captures the balance between the tendency to keep and honor past traditions, and the drive for pragmatic, future-oriented problem solving. This cultural dimension  might be partly reflected by the historical times the honorees represent---whether they represent a distant or a closer past. The history of the four cities has been rich of major societal and cultural revolutions that spanned the whole two past centuries, and that have influenced contemporary culture to different extents. We looked at how the protagonists of these past events are celebrated through street names.

\mbox{ }\\
\noindent
\textit{\textbf{Q3} What are the most celebrated professions?}\\
The way in which the honorees' professions are celebrated might give an indication about the values that are associated to personal status. Societies characterized by higher levels of power distance are more inclined to build highly hierarchical social structures and to celebrate high-status individuals---for example, those who have elite professions. The occupation might also contain information about the types of achievements that are most celebrated, which ties back to the notion of individualism: individualistic societies value more personal achievements (e.g., a sportsman breaking a world record) rather than collective achievements (e.g., a civil right activist). 
Avenues of Washington DC are named after the member states of the Union~\cite{azaryahu1996power}, while streets in Bucharest celebrate the working class~\cite{light2004street}. Similarly, we explored the professions of the individuals commemorated in the streets of the four cities as a way to determine what these societies valued over the years.   

\mbox{ }\\
\noindent
\textit{\textbf{Q4} Are foreigners celebrated?} \\
Collectivist societies, as opposed to individualistic ones, tend to be more patriotic~\cite{house2004culture,taylor2012does}: street naming is often an expression of power and patriotism, and the majority of streets are named after local and national leaders. But, at times, streets are named after foreigners as well. The extent to which this happens is largely determined by historical events (alliances, colonization, or immigration), and by the position of society in the spectrum of patriotism. Therefore, we looked at the extent to which street names celebrate foreign cultures.

\section*{METHODS}
\label{sec:section4}

\begin{table}
\caption{Dataset Statistics. Each city is listed with its streets' denomination periods and the historical periods in which the corresponding honorees lived. These are shown as (min, max) ranges.}
\label{tab:stats}
\begin{tabular}{llll}
\hline
City & Honorific Streets & Denomination Period & Historical Period Honorees \\
\hline
Paris & 1428 & 1202 - 2011 & 60 B.C. - 1940 \\
Vienna & 1662 & 1778 - 2018 & 700 - 1982\\
London & 770 & 1030 - 2013 &  18 B.C. - 1961 \\
New York & 1072 & 1998 - 2013 & 1474 - 2003\\ \\\hline
\end{tabular}
\end{table}

\subsection*{Datasets}
We gathered street name information about the four cities from open data sources. For each street, we collected eight types of information: the district that the street belongs to, the renaming date (denomination) of the street, the person name (honoree) that the street was named after, the honoree's gender, the honoree's occupation, the honoree's country of origin, the honoree's date of birth (dob), and date of death (dod). These curated datasets include only information about honorific streets, namely streets that are dedicated to historical figures --- this is a subset of the total set of streets. In this study, we focus on honorific streets and we do not consider streets that are not named after people (e.g., places, events, or natural elements). Next, we describe how we obtained, cleaned, and structured this data (Table~\ref{tab:stats}), which we made publicly available (\url{https://social-dynamics.net/streetonomics/data}) to support future research.

\mbox{} \\
\noindent
\emph{Paris.}
We used both Wikidata and Wikipedia. We queried the structured database Wikidata through its SPARQL endpoint query service, and obtained all the instances of the ``street class'' located in Paris, which amounted to 3,413 streets. Each returned street object had a ``named after'' field, which was a link to the corresponding eponym. Streets in Paris are named mostly after regions, events, or people. Since we were interested in streets named after people, we took all eponyms instances of the ``person class'' (leaving us with 1,808 streets). We then crawled these  persons' Wikipedia pages, and that resulted in a dataset in which each row contained a street name; its borough; and its honoree's name, gender, occupation, date of birth, date of death, and country of origin. After further data cleaning, we ended up with the final Paris dataset containing information for 1,428 honorific streets.

\mbox{} \\
\noindent
\emph{Vienna.}
We started from the `Vienna History Wiki' platform (\url{https://www.geschichtewiki.wien.gv.at}), which brings together historical knowledge about the city, and it does so in structured webpages. Each webpage contains the history of a given street and an infoBox (similar to Wikipedia) containing street creation date, denomination (if the street was named after a person), and district. We initially resolved data about 2,481 streets. We translated German names to English in two steps: first, we used an off-the-shelf python package for language translation using Google Translate API (\url{https://pypi.org/project/googletrans}). Second, to resolve any disambiguation from the automatic translation, one of the authors manually corrected 5\% of them (e.g., disambiguation due to special characters in the German language). After data cleaning (i.e., streets whose honoree full information could not be found), we ended up with the final Vienna dataset containing 1,662 streets.

\mbox{} \\
\noindent
\emph{London.}
We crawled Wikipedia entries that referred to London streets, narrowing down the search on  entries that contained mentions that typically characterize pages about streets such as  `named after', `honor', and `celebrate'. This process resulted into 2500 streets, which were then annotated using the Amazon Mechanical Turk crowd-sourcing platform: each entry was annotated by at least 2 workers, and conflicts were manually resolved. The resulting dataset consisted of 770 streets.

\mbox{} \\
\noindent
\emph{New York.}
From urban planner Gilbert Tauber (\url{http://nycstreets.info}), we obtained a curated dataset that contains 1,459 honorific streets, and each street is associated with its honoree name and gender, and the street's denomination and borough. After data cleaning, we ended up with the final New York dataset containing 1,072 streets.

\mbox{ } \\ \mbox{ }
We then mapped all this information into each city's neighborhoods defined by the official polygons.

\begin{figure*}
    \centering 
    \captionsetup[subfigure]{justification=centering}
\begin{subfigure}{\textwidth}
  \includegraphics[width=\linewidth]{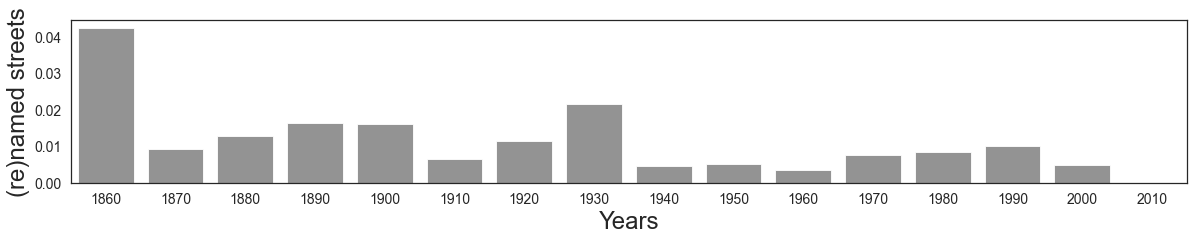}
  \caption{Paris}
  \label{fig:renaming_1}
\end{subfigure}\hfil 
\begin{subfigure}{\textwidth}
  \includegraphics[width=\linewidth]{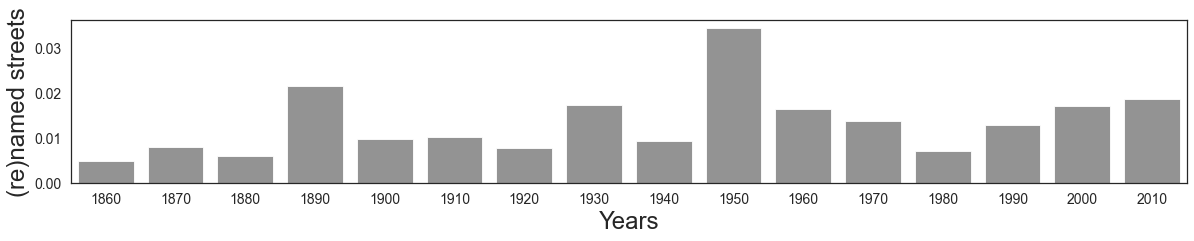}
  \caption{Vienna}
  \label{fig:renaming_2}
\end{subfigure}\hfil 
\begin{subfigure}{\textwidth}
  \includegraphics[width=\linewidth]{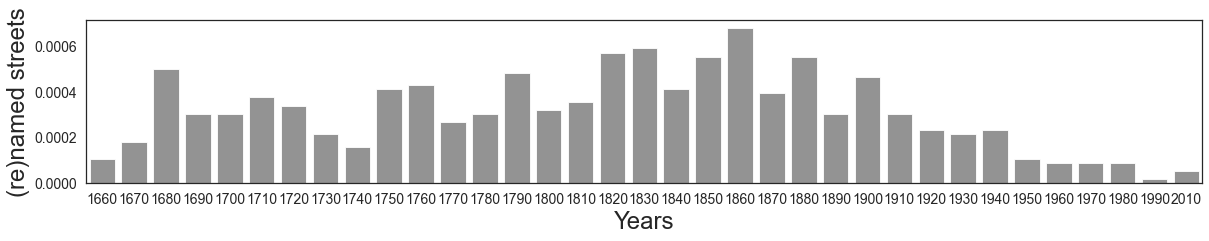}
  \caption{London}
  \label{fig:renaming_3}
\end{subfigure}\hfil
\begin{subfigure}{\textwidth}
  \includegraphics[width=\linewidth]{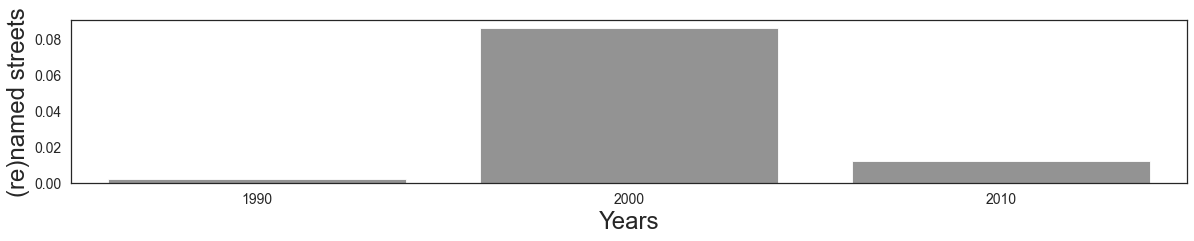}
  \caption{New York}
  \label{fig:renaming_4}
\end{subfigure}
\caption{The fraction of streets existing today that were (re)named over the decades. Results are shown \newline by city.}
\label{fig:street_renaming}
\end{figure*}

\begin{figure*}
    \centering 
\begin{subfigure}{0.24\textwidth}
  \includegraphics[width=\linewidth]{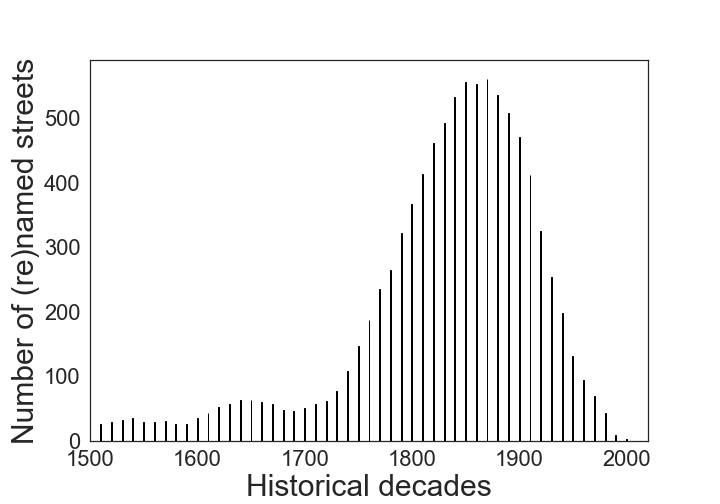}
  \caption{Paris}
  \label{fig:dob_1}
\end{subfigure}\hfil 
\begin{subfigure}{0.24\textwidth}
  \includegraphics[width=\linewidth]{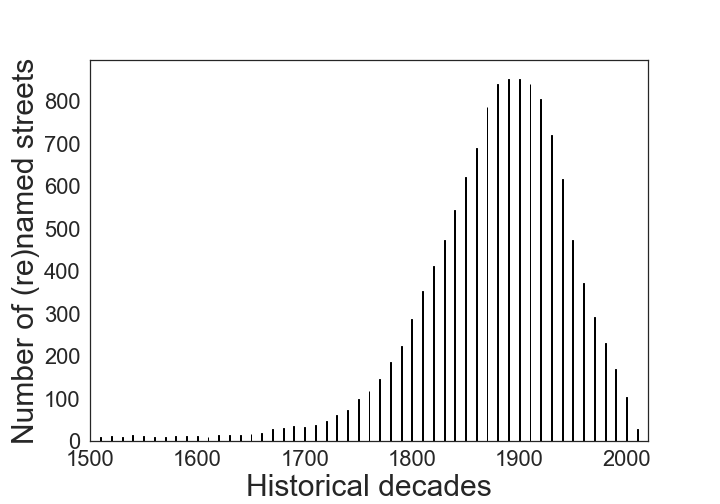}
  \caption{Vienna}
  \label{fig:dob_2}
\end{subfigure}\hfil 
\begin{subfigure}{0.24\textwidth}
  \includegraphics[width=\linewidth]{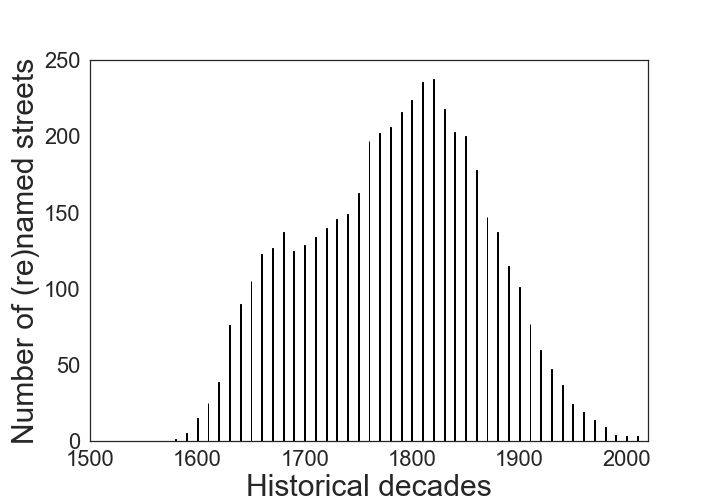}
  \caption{London}
  \label{fig:dob_3}
\end{subfigure}\hfil 
\begin{subfigure}{0.24\textwidth}
  \includegraphics[width=\linewidth]{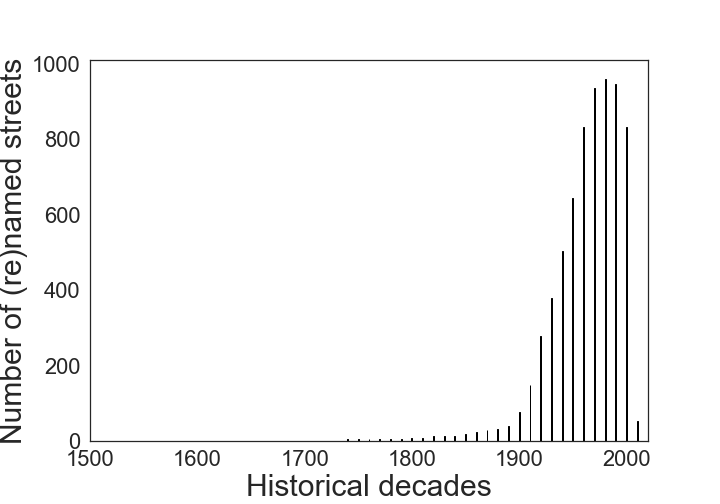}
  \caption{New York}
  \label{fig:dob_4}
\end{subfigure}
\caption{Number of honorees (after whom all the city's streets were named) \newline who were alive at each historical decade.}
\label{fig:dob_results}
\end{figure*}

\subsection*{Data Processing and Formulae}
Having obtained these four datasets, we processed them as follows.
\mbox{}
\newline
\newline
\noindent
\emph{Temporal and spatial mapping.} We mapped our data spatially in terms of district \emph{D} (neighborhood); and temporally in terms of two types of decade: denomination decade \emph{DD}  (the decade in which a street was denominated), and focus on historical decade \emph{FHD} (the extent to which an historical  decade is represented across a city's streets, which is based on the decades during which all the honorees lived).

\mbox{} \\
\noindent
\emph{Focus on Historical Decade (FHD).} We used an honoree's dob and dod to reflect his/her lifespan. We then counted the number of times the honoree's lifespan coincides with each historical decade. We repeated that for all the honorees (i.e., for all the city's streets in our data). The resulting measure---the (city) focus on historical decade---reflects the extent to which each decade is represented across the city's streets. \vspace{1mm}

\mbox{} \\
\noindent
\emph{Gender.} We computed a female proportion (\emph{f\_prop}) per denomination decade DD and per district D. The female proportion per D is computed by first grouping streets associated with females into their districts, and by then counting the proportion of these streets in each district \textit{D}. 
\begin{equation}
f\_prop@DD=\frac{\#f\_streets@DD}{\#all\_streets@DD}
\label{eq:females_proportion_denomination}
\end{equation}
\begin{equation}
  f\_prop@D = \frac{\#f\_ streets@D}{\#all\_ streets}
\label{eq:females_proportion_district}
\end{equation}

\mbox{} \\
\noindent
\emph{Foreigners.} After marking as \emph{foreigners} the honorees whose countries of origin were different than the city's country, we computed the proportion of foreigners (\emph{for\_prop}) per denomination decade DD and per district D: 
\begin{equation}
  for\_prop@DD = \frac{\#for\_ streets}{\#all\_ streets@DD}
\label{eq:foreigners_proportion_denomination}
\end{equation}
\begin{equation}
  for\_prop@D = \frac{\#for\_ streets@D}{\#all\_ streets}
\label{eq:foreigners_proportion_district}
\end{equation}

\mbox{} \\
\noindent
\emph{Occupation.} Based on the International Standard Classification of Occupations, we mapped each occupation to any of the following 12 occupational groups:  creative and performing artists, authors, journalists and linguists, science and engineering professionals, legal, social and cultural professionals, craft and related trades workers, business and administration professionals, legislators, commissioned armed forces officers, religious, health associate professionals, and teaching professionals. This category-based classification schema allowed us to map historical professions and modern professions under a unified set of categories. For example, bishops, archibishops, or clergy were mapped into the `religous' category; in a similar way, we mapped the rest of the professions' categories.

\subsubsection*{Deciding the starting date.}
We set the starting decade for Paris to 1860, which is when the city began its expansion and became what it is  nowadays.  The number of streets (re)named peaked in the 1920-30s (Figure~\ref{fig:street_renaming}a). That period is the post-World War I era during which writers and artists from all over the world were attracted to Paris.  For Vienna, our analysis started in 1860, the time when the modern city was re-purposed. After the World War II (Figure~\ref{fig:street_renaming}b), the city underwent a large-scale urban development, and streets were named after key figures of the war. In London, we started in 1666, when the Great Fire broke out and the city was reborn from its ashes (Figure~\ref{fig:street_renaming}c). In New York, we started our analysis in 1998, when the street co-naming was introduced. We observed that there was a drastic increase in the number of streets being named after the 9/11 terrorist attack (Figure~\ref{fig:street_renaming}d), where hundreds of streets were named in memory of victims, emergency responders, and people in the community who provided help. 

\section*{RESULTS}
\label{sec:section5}

\begin{figure*}
  \centering 
\begin{subfigure}{0.48\textwidth}
\includegraphics[width=\linewidth]{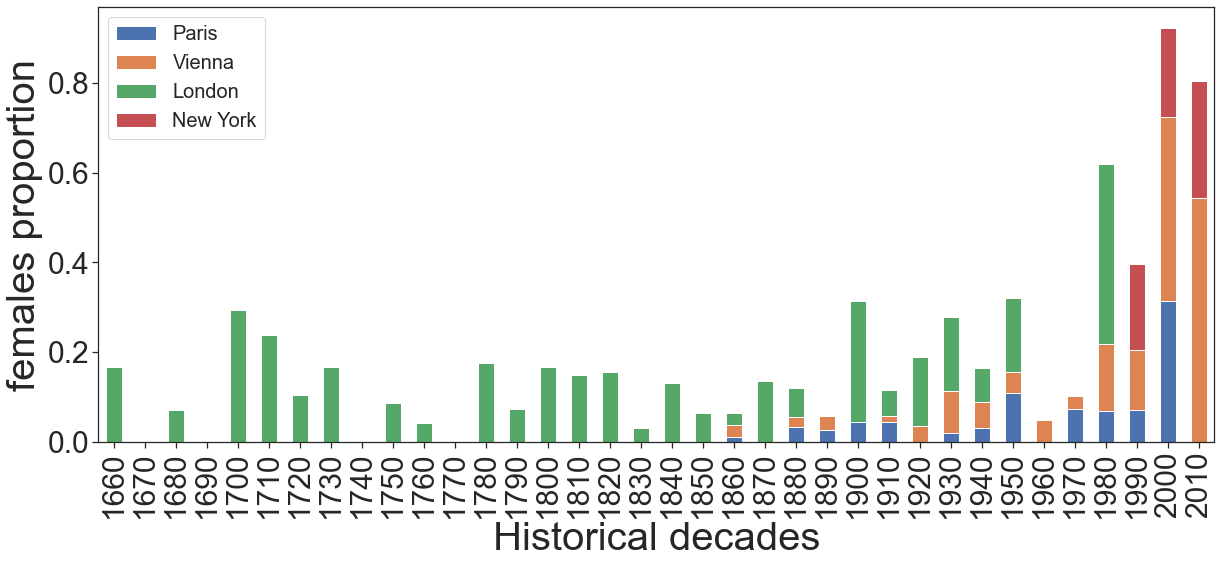}
  \caption{streets named after females}
  \label{fig:gender_all}
\end{subfigure}\hfil 
\begin{subfigure}{0.48\textwidth}
  \includegraphics[width=\linewidth]{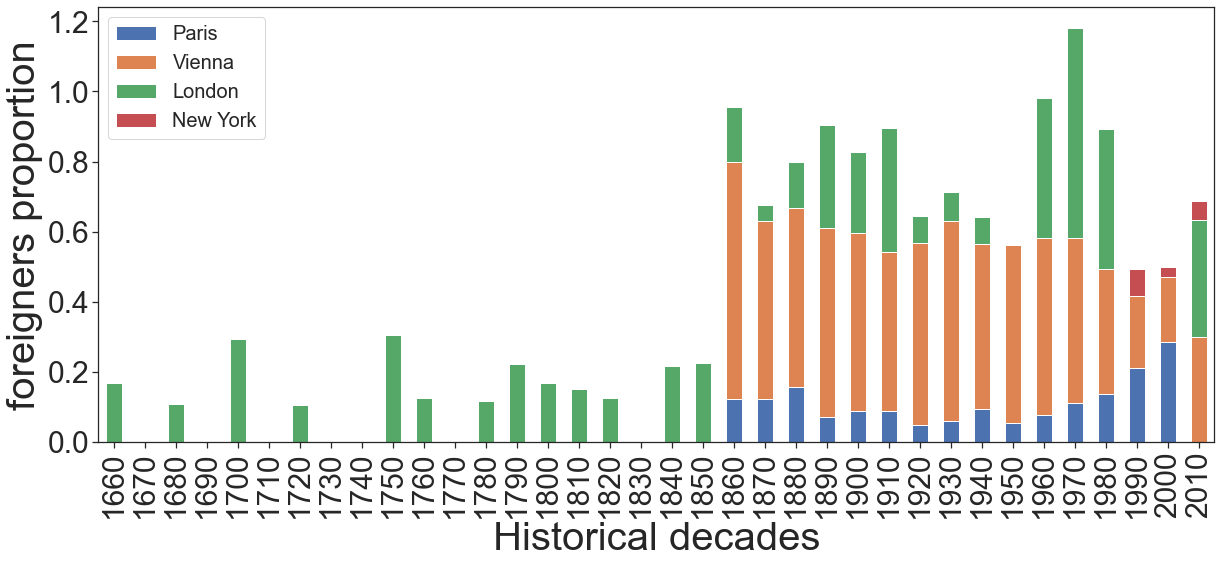}
  \caption{streets named after foreigners}
  \label{fig:foreigners_all}
\end{subfigure}\hfil 

\caption{Over decades, the proportion of streets named after: (a) female \newline figures using Formula (\ref{eq:females_proportion_denomination}), and  (b) foreigners using Formula (\ref{eq:foreigners_proportion_denomination}). Historical decades refer to the date of (re)naming of a street.}

\label{fig:gender_foreigners_temporal}
\end{figure*}

\begin{figure*}
    \centering 
\begin{subfigure}{0.85\textwidth}
  \includegraphics[width=\linewidth]{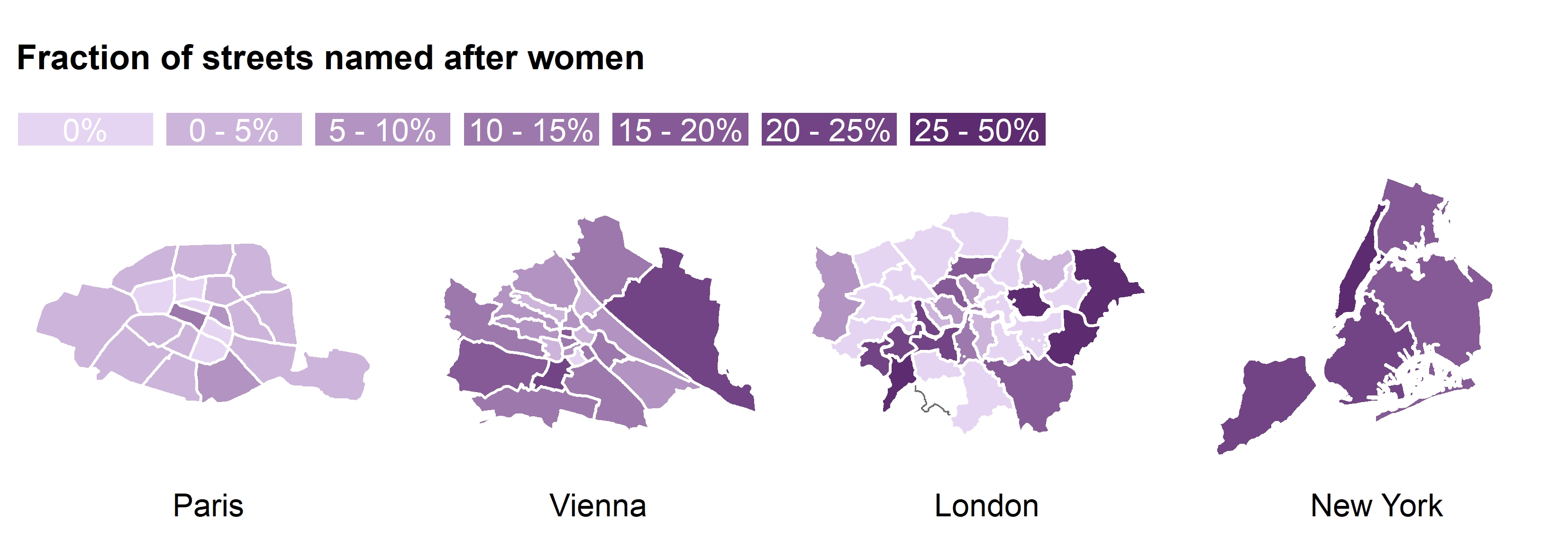}
  \caption{}
  \label{fig:gender_new}
\end{subfigure}\hfil 
\begin{subfigure}{0.85\textwidth}
  \includegraphics[width=\linewidth]{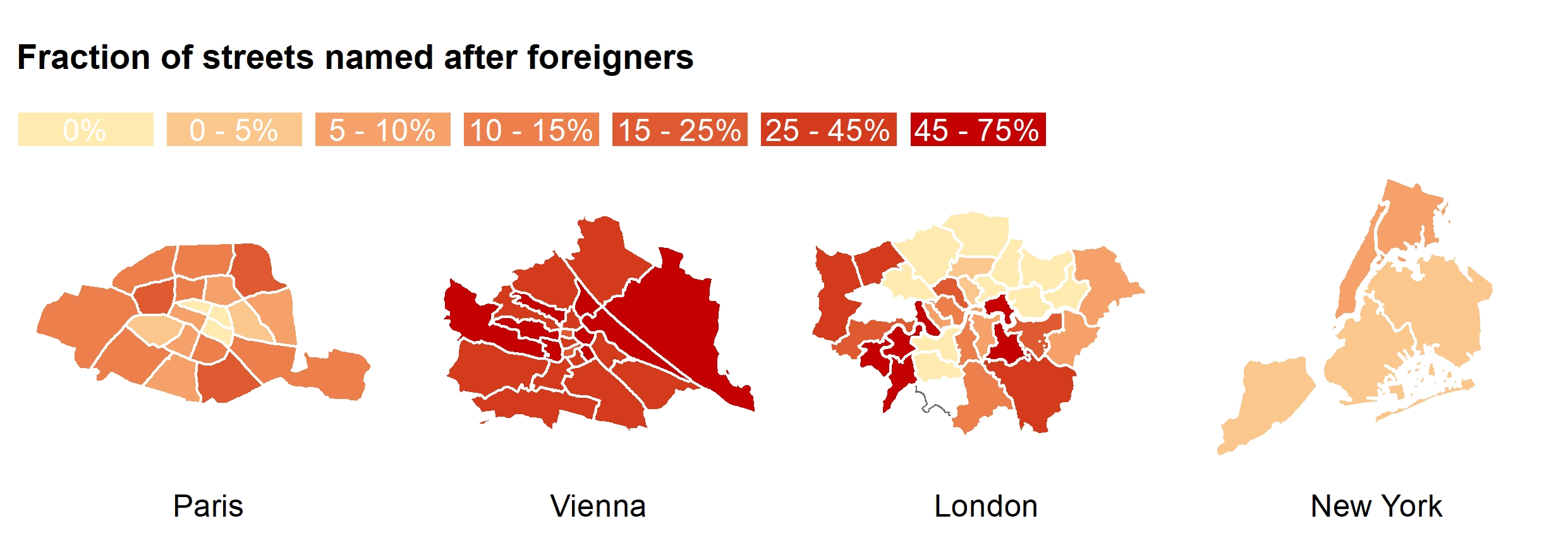}
  \caption{}
  \label{fig:foreigners_new}
\end{subfigure}\hfil 
\caption{Fraction of streets named after: \emph{(top)} (a) women computed \newline with formula (\ref{eq:females_proportion_district}); and \emph{(bottom)} (b) foreigners computed with formula (\ref{eq:foreigners_proportion_district}). \newline The more intense the color, the higher the fraction.}

\label{fig:gender_foreigners_version2}
\end{figure*}

\subsection*{Gender}
\label{sec:gender}
\setlength\itemsep{1em}
\noindent
\textit{\textbf{Q1} Are streets names in Paris, Vienna, London, and New York gender-biased?}\\
\setlength\itemsep{1em}
Despite Paris and New York being centers of progressive culture, they still fail to commemorate men and women in equal measure. Sadly, what is an historically legacy will stay with us in the future, if these two cities do not change their course.  As men's and women's aspiration converge, street naming should reflect that. More specifically, in Paris, 4\% of today's streets are named after women, and these streets are clustered in the 1\textsuperscript{st} arrondissement, the very center of Paris (Figure~\ref{fig:gender_foreigners_version2}a), and that proportion used to be  only 1\% prior to 1980 (Figure~\ref{fig:gender_foreigners_temporal}a). In the 1880s, in Vienna (Figure~\ref{fig:gender_foreigners_temporal}a),  the proportion of streets named after females was below 5\%, reaching a peak of 10\% in 1920-1940s. In 1980s, more and more streets started being named and renamed after female figures, resulting into a proportion of female streets as high as 41\% in 2000s and 54\% in 2010s. These streets are predominantly clustered in the 22\textsuperscript{nd} district (Donaustadt), in the east side of the city (Figure~\ref{fig:gender_foreigners_version2}a). In London, the proportion of female steets reached its highest proportion of 40\% around the 1980s (Figure~\ref{fig:gender_foreigners_temporal}a), and are predominantly clustered in the South West boroughs. Finally, in New York, the proportion of female streets rose steadily over the last three decades, up to 26\% in 2010s (Figure~\ref{fig:gender_foreigners_temporal}a), and are clustered in Manhattan (Figure~\ref{fig:gender_foreigners_version2}a).  

\subsection*{Historical decades}
\setlength\itemsep{1em}
\noindent
\textit{\textbf{Q2} Are the historical times in which the honorees lived close or distant from present times?}\\
\setlength\itemsep{1em}
Streets in Paris are named mostly after people who lived in the 1860s, as the peaks in the two distributions (Figure ~\ref{fig:street_renaming}a and ~\ref{fig:dob_results}a) suggest approximately the same time period. This reflects the era of the Second Empire during which urban planner Haussmann worked with Napoleon III to transform Paris into an empire city. Form comes with function, and Haussmann's vision transformed the city both spatially and socially. His urban changes entailed better infrastructure, including the reconstruction of the entire Paris's sewage system~\cite{gandy1999paris}, and left a mark that is still visible today~\cite{de2002paris}.

In Vienna, most of the historical figures were living through the 1900s (Figure~\ref{fig:dob_results}b). Around this time, the city had expanded. The historical period most represented in the city's streets appear to be that between World War I and the Nazi invasion~\cite{olsen1986city}. World War I destroyed the city and, around 1940s, Austria was invaded by Nazi Germany. 

London streets are named mostly after people who lived through the 1700s and 1800s. Within that temporal window, the city experienced the Great Fire, grew because of the large-scale interventions promoted by King George III, and benefited from the vibrant atmosphere created by the introduction of the printing press.

Finally, in New York, streets celebrate people who lived through 1950s to 2000s (Figure~\ref{fig:dob_results}c), and a considerable part of them (36\%) are named after 9/11 victims and emergency responders.

\begin{figure*}
\centerline{
\includegraphics[width=\linewidth]{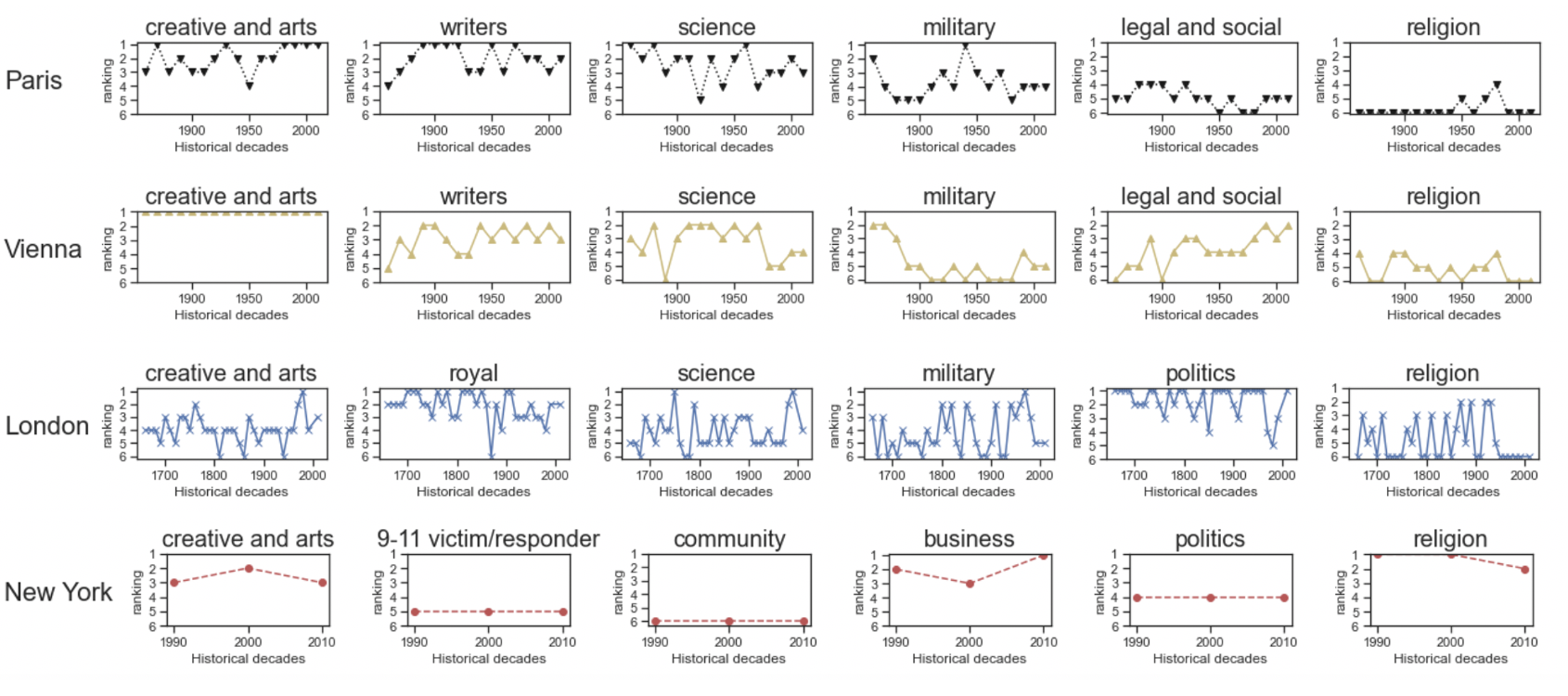}
}
\caption{The rise and fall of occupations. Each occupation is ranked by \newline its overall frequency among streets at each decade.}
\label{fig:occupations_ranking}
\end{figure*}

\subsection*{Occupation}
\setlength\itemsep{1em}
\noindent
\textit{\textbf{Q3} What are the most celebrated professions?}\\
\setlength\itemsep{1em}
We analyzed the professions of the individuals commemorated in the streetscapes of Paris, Vienna, London, and New York to determine what these societies valued over the decades.

As Paris's value system changed, the importance of professions changed as well. After World War II, most streets commemorated people in the military (ranked 1\textsuperscript{st} during 1940-1950s). In the decades that followed, Paris was reborn from its ashes and honored more artists, writers and scientists, leading the military profession to be ranked at the 5\textsuperscript{th} position nowadays. Religious occupations did not received much attention, except the period between 1950s and 1980s, during which they ranked at the 4\textsuperscript{th} position. 

Vienna's  streets celebrate artists (Figure~\ref{fig:occupations_ranking}a). Yet its streets still reflect historical changes in society as it has been in the case of Paris. People in the military ranked at the 3\textsuperscript{rd} position in 1860s (after the Austria-Hungary war), increasingly lost importance over time, and finally ended up at the 9\textsuperscript{th} position today. Compared to Paris, scientists and writers have not been valued as much, while those in legal and social occupations have  (they are at the 3\textsuperscript{th} position today).

London's streets celebrate the British royal family, politicians, and military professionals. Artists, on the other hand, were neglected for centuries and have only recently been represented, reflecting the more contemporary image of the city.

In New York, the number of honorific street names peaked in 2001 after the 9/11 attack, and these streets today represent 36\% of the total  (Figure~\ref{fig:street_renaming}c). This has resulted in the celebration of people who worked in the `business and community'  area (e.g., civil servants). The profession that has remained constant is that of artists, which have constantly ranked at the  4\textsuperscript{th} position.

To study the long-term naming trends, we grouped all the decades into the four half-centuries of 1800-1850, 1850-1900, 1900-1950, and 1950-2000. For each of the half-centuries, we computed the average of its five correlation coefficients; each coefficient is the Kendall Tau coefficient between the occupational ranking in a given decade (e.g., 1850) and that of the next (e.g., 1860). We did so for all the cities but New York. That is because its street renaming patterns have been comparatively recent and, as such, they can be studied one decade over another but are not amenable to be studied over the two centuries.

\begin{figure*}
\centerline{
\includegraphics[width=0.8\linewidth]{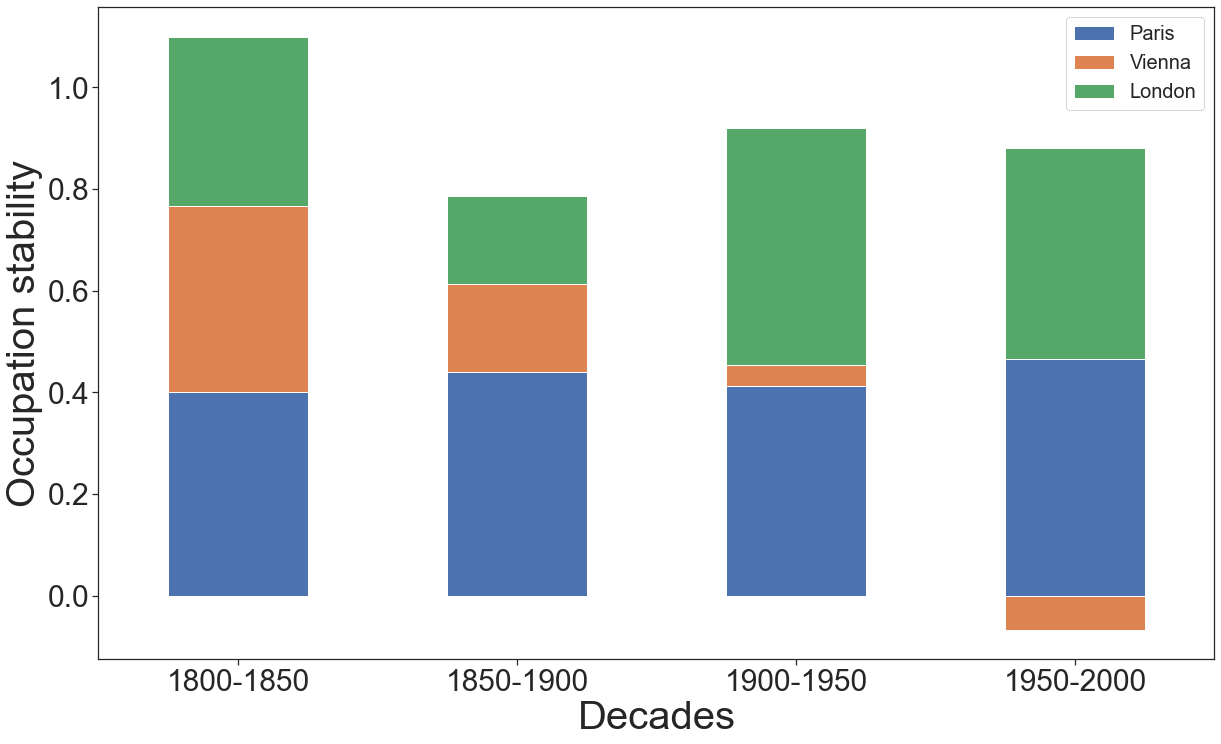}
}
\caption{The stability of the types of occupations celebrated by a city's streets. The higher a city's value, the more stable the occupations celebrated by its streets remained in a given half-century.}
\label{fig:occupations_kendall_tau}
\end{figure*}

Figure~\ref{fig:occupations_kendall_tau} shows the average correlations for all the four half-centuries.
For a given half-century, the lower the correlation, the more changes the city experienced in terms of celebrated occupations. We see that  Paris  experienced incremental and constant changes throughout the centuries, Vienna mostly in the $20^{th}$ century (which followed the biggest expansion started with the Ringstrasse era), and London during the first half of that century (during which the city experienced its largest expansion).

\subsection*{Foreigners}
\setlength\itemsep{1em}
\noindent
\textit{\textbf{Q4} Are foreigners celebrated?} \\
\setlength\itemsep{1em}
The overall proportion of streets named after foreigners (Figure~\ref{fig:gender_foreigners_temporal}b) is highest in Vienna (44.6\%), followed by London (14.6\%), Paris (10.9\%), and then New York (3.2\%).

In Paris, the great majority of streets named after foreigners are clustered in the 8\textsuperscript{th} arrondissment (Figure~\ref{fig:gender_foreigners_version2}b), in the northwest part of the city. At the beginning of great expansion (1860-1880), about 18\% of the streets were named after foreigners (Figure~\ref{fig:gender_foreigners_temporal}b). That was after the French revolution (1785) and, at that point, Belgium was part of France. After the Belgian revolution (1830-1839), the two nations split, and the number of streets named after foreigners decreased to 10\%, a proportion that lasted for a century. Quite recently (1980s), the proportion rose again to reach 29\% in 2000s. Interestingly, in the last decade, there was not any new street (re)named after a foreigner.

As for Vienna, 44.6\% of its streets are named after foreigners (Figure~\ref{fig:gender_foreigners_temporal}b), reflecting the city's historical ties with other Central and Eastern European countries. The proportion of foreigners in street names peaked in the 1870s (70\%), right after the formation of the Austrian-Hungary republic. After 1900s, the proportion  started to decrease, down to 45\% in 1910s, and to 18\% during 2000s. As opposed to Paris, in the last decade, the proportion rose again and reached 30\% in present times. As one expects, most of these streets  are located in the center of the city (Figure~\ref{fig:gender_foreigners_version2}b), that is, in the inner old town (now Innere Stadt, 1\textsuperscript{st} district), and in the northern part (Brigittenau, the  20\textsuperscript{th} district).

London streets celebrate foreigners less than what Vienna does, but more than what the two other  cities do (Figure~\ref{fig:gender_foreigners_temporal}b). A decline in the number of streets named after foreigners was experienced  shortly after the two world wars, suggesting a more inward-looking perspective. There was then a reversal of fortune, with a peak being registered in the 1970s, period in which the city financially and culturally took center stage.

The case of New York is different. Only 5\% of its streets are named after foreigners (Figure~\ref{fig:gender_foreigners_temporal}b), clustered mostly in the Bronx (Figure~\ref{fig:gender_foreigners_version2}b). The proportion was only marginally higher (7\%) two decades back. New York is the very model of a global city, but it is more so financially than culturally. That might be partly because the city is quite young compared to Vienna or Paris, and partly because it tends to focus on itself. New York is the world's financial capital. Yet art, music and fashion have contributed to the city's economy for a long time, but their contribution has been overlooked. This cultural production ``relies more than ever on the social encounters that come from living shoulder to shoulder with your peers'' (\url{https://www.economist.com/books-and-arts/2007/09/06/art-and-the-city}). Since a considerable number of these peers are foreigners, we expect that New York will change its course and start to celebrate some of these individuals, in line with its truly global nature.

\begin{table}
\caption{Validity check.}
\label{tab:stats_validity}
\begin{tabular}{llllll}
\hline
City & \shortstack[c]{Total streets \\(from OSM)} & \shortstack[c]{People (\%)\\(our dataset)} & \shortstack[c]{People (\%)\\(200 streets)} & \shortstack[c]{Female (\%)\\(our dataset)} & \shortstack[c]{Female (\%)\\(sample)} \\
\hline
Paris & 6953 & 21 & 46 & 6 & 4\\
Vienna & 7498 & 22 & 56 & 12 & 11\\
London & 55921 & 1 & 30 & 10 & 24\\
New York & 10438 & 10 & 14 & 20 & 14\\ \\\hline
\end{tabular}
\end{table}

\subsection*{Validation}
To ensure the robustness of our findings, we conducted an additional validity check, with the main goal of estimating two quantities: i) the coverage of our curated datasets over the total set of honorific streets existing in the four cities, and ii) the actual male/female ratio in the full set of honorific streets, as a simple way to appraise the error of the measures we used in our study.

Tracing back the name of a street to its honoree is a demanding task, and it requires the investigation of multiple historical records that are often not available in digital format. Attempting to do so without relying on such historical sources faces several challenges including that of name expansion (i.e., inferring the honoree's first name from the last name) and disambiguation (i.e., when a street name could refer to multiple entities, such as a place or a person). With these challenges in mind, we adopted a best-effort approach to obtain a rough estimate of the two target quantities on a random set of streets.

We used a two-step methodology. First, we obtained road segments within administrative borders of each city from OpenStreetMap (\url{http://download.geofabrik.de}). The dataset includes road segments of different types such as residential roads or pedestrian tracks (\url{https://wiki.openstreetmap.org/wiki/Key:highway}). We excluded motorways, trunks, cycleways and paths in addition to streets that included numbers or no name was attached to them. This yielded a total number of 6,953 in Paris, 7,498 in Vienna, 55,921 in London, and 10,438 in New York. 
Second, for each city, we randomly sampled 200 streets uniformly distributed across the city's boroughs/districts and searched those street names on Wikipedia. We marked a street as a honorific if, among all the search results, there was an explicit mention about its history. For all honorific streets obtained as such, we obtained the gender of the honoree (Table~\ref{tab:stats_validity}). In the city of Paris, we found a total of 92 streets linked with people, whereas the rest 108 are not honorific. Out of the 46\% of honorific streets, we found a total of 6\% to be named after a female figure. Similarly, in Vienna, we found 56\% streets in the random sample to be honorific, and out of these 11\% were name after women. In London, we found 30\% out of the 200 streets to be honorific, and out of these a total of 10\% were named after women. In New York, we found 14\% out of the 200 streets to be honorific, and out of these a total of 14\% were named after women. This additional validity check illustrates that, to a great extent, our findings are aligned with a smaller random sample manually checked.

\section*{Discussion and conclusion}
\label{sec:section6}

\subsection*{Main Results}
By analyzing honorific streets (i.e., streets that were named after a person), we revealed social-cultural characteristics of the four cities under study. Our findings, however, refer to a subset of streets (i.e., honorific streets) in those cities, and they should be interpreted as such. We found that streets in Paris, Vienna, London and New York are gender-biased. Such a bias has been mitigated by a recent trend of naming new streets after outstanding female figures. Yet, Paris, for example, despite being at the center of gender equality movements and public debates, still remains the most gender-biased city among the three. 

As one expects, street names reflect a country's culture and can be used as proxies for the country's historical and cultural characteristics (e.g., as an indicator of religiosity~\cite{oto2017street}). New York's street naming is more present-oriented than Vienna's or Paris's or London's. The streets of the three European capitals are mostly named after people who lived in the 19\textsuperscript{th} century, while those in New York, even when not considering the 9/11 victims as part of the picture, are mostly named after people who had spent significant part of their lives in the second half of the 20\textsuperscript{th} century.

The representation of occupations over time shows an interesting evolution. Military professions were celebrated after major conflicts, and then declined as the western world entered in its longest period of peace after World War II. The streetscape in Paris (and all the more so in Vienna) have always been dominated by artists or writers, and that perfectly reflects the role of these two cities as the world's avant-garde for arts, fashion, and beauty in the 20\textsuperscript{th} century. Interestingly, streets named after high-status professions such as lawyers and jurists declined over time in Austria, while they increased in France. As one expects, London tend to celebrate royals. 
In New York, most of the streets are named after people who worked in the `business and community' area (e.g., civil servants).

Finally, Vienna has more than 44.6\% of streets named after foreigners. This mainly reflects the city's historical ties with Central and Eastern European countries other than Austria (e.g., with Hungary through the creation of the Austrian-Hungary Empire of dual monarchy in 1870s, with Germany through the invasion of Austrian-born Adolf Hitler in 1938). London ranked second among the four cities with a fraction of 14.6\% of its streets named after foreigners. Paris has a small fraction of its streets named after foreigners (10.9\%), and the Belgian's influence on those streets is easily explained: Belgium and France had been one country for a short while. Finally, despite being considered a global city, New York tends to be self-focused by  almost exclusively celebrating Americans (only 3.2\% of its streets are named after foreigners).

\subsection*{Implications}
Our methodology offers an alternative way to study urban culture through open data. Cultural studies have been typically conducted through the manual analysis of literary production. More recently, they have been conducted at scale through the analysis of millions of digitized books~\cite{michel2011quantitative}. In line with previous work, we have shown that also street names are indicators of culture  because they encapsulate a society's beliefs and, consequently, their analysis could be used to study hypotheses that have been left hitherto untested, and these hypotheses could be taken from different fields, from Urban Studies to Digital Humanities.
 
From a practical standpoint, our work might benefit two main stakeholders. First, public authorities and municipalities could reflect on their past to inform their future. To see how, consider the city of Madrid. To mark its distance from the dictatorship period, it has recently started to change the street names that were assigned during that period. Our tools can stimulate this type of public reflection by enabling policy makers to take well-informed decisions, and for citizens to participate in discussions. Ultimately, the idea behind these tools is to increase historical awareness, promote cultural heritage, and reshape the collective memory positively. Second, as people tend to forget their historical past~\cite{ebbinghaus2013memory, halbwachs1992collective}, our methodology could help raise dwellers' cultural and historical awareness. Educational tools in the form of simple games or visual analytics might well improve people’s awareness of their city's history and enable new forms of grass-root engagement~\cite{edyta20}. As a case in point, consider a mobile app that monitors its user's mobility, quantify the historical bits the user is exposed to on a daily basis, and returns summaries of key historical events  back on a weekly basis.

\subsection*{Limitations and future work}
This study has four limitations that call for further work. First, our methodology is not fully automatic. The procedure for obtaining and creating structured datasets from heterogeneous data sources is automatized, but cases in which data was not present or badly formatted were manually resolved. As such, this study cannot be easily scaled to the entire world. Language translation, data collection, and data integration are just some of the technical challenges to be solved. This leads us to our second limitation: our analysis is restricted to three cities located in Europe, and one in the United States. This set contains world-class cities that are excellent candidates to study the evolution of culture in the Western world through centuries, but it does not represent other blocks of cultural or religious identities (e.g., Islamic world, Latin-American cultures). The third limitation concerns potential selection biases in the open data sources used in the study. A case in point is women. They are represented in Wikipedia pages less than men~\cite{wagner2015s}. Future work could explore the extent to which the street naming biases we reported are further aggravated by the biases of the external data sources we relied on. Finally, the fourth limitation is that our analysis is limited to an important yet tiny portion of history: the last 200 years, and three decades in the case of New York (during that time honorific streets were introduced in New York) Yet the very same methodology could be applied in different historical periods, which could lead to additional cross-cultural comparisons.

\nolinenumbers

%
%
%

\end{document}